\documentclass[12pt, letterpaper]{article}
\pdfoutput=1
\usepackage[pdftex]{graphicx}
\usepackage{epsfig}
\usepackage{latexsym}
\usepackage{amssymb}
\usepackage{amsmath}

\usepackage{wrapfig}
\usepackage{boxedminipage}
\usepackage{setspace}
\usepackage{subfigure}

\DeclareSymbolFont{AMSb}{U}{msb}{m}{n}
\DeclareMathSymbol{\IN}{\mathbin}{AMSb}{"4E}
\DeclareMathSymbol{\IZ}{\mathbin}{AMSb}{"5A}
\DeclareMathSymbol{\IR}{\mathbin}{AMSb}{"52}
\DeclareMathSymbol{\Q}{\mathbin}{AMSb}{"51}
\DeclareMathSymbol{\II}{\mathbin}{AMSb}{"49}
\DeclareMathSymbol{\IC}{\mathbin}{AMSb}{"43}
\DeclareMathSymbol{\IP}{\mathbin}{AMSb}{"50}
\DeclareMathSymbol{\IH}{\mathbin}{AMSb}{"48}
\DeclareMathSymbol\IA{\mathalpha}{AMSb}{"41}
\DeclareMathSymbol\IS{\mathalpha}{AMSb}{"53}

\def\Q{{\cal Q}}

\begin{document}

\begin{flushright}
\phantom{{\tt arXiv:0802.????}}
\end{flushright}

\bigskip
\bigskip
\bigskip

\begin{center} {\Large \bf  Universality in the Large $N_c$ Dynamics of Flavour:}

\bigskip

{\Large\bf  Thermal {\sl Vs.}  Quantum Induced Phase Transitions}

\end{center}

\bigskip \bigskip \bigskip \bigskip

\centerline{\bf Veselin G. Filev, Clifford V. Johnson}

\bigskip

\bigskip
\bigskip

  \centerline{\it Department of Physics and Astronomy }
\centerline{\it University of
Southern California}
\centerline{\it Los Angeles, CA 90089-0484, U.S.A.}

\bigskip

\centerline{\small \tt  filev, johnson1 [at] usc.edu}

\bigskip
\bigskip


\begin{abstract} 
\noindent

We show how two important types of phase transition in large $N_c$
gauge theory with fundamental flavours can be cast into the same
classifying framework as the meson--melting phase transition. These
are quantum fluctuation induced transitions in the presence of an
external electric field, or a chemical potential for R--charge. The
classifying framework involves the study of the local geometry of a
special D--brane embedding, which seeds a self--similar spiral
structure in the space of embeddings. The properties of this spiral,
characterized by a pair of numbers, capture some key universal
features of the transition. Computing these numbers for these
 non--thermal cases, we find that these transitions are in
the same universality class as each other, but have different
universal features from the thermal case.  We present a natural
generalization that yields new universality classes that may pertain
to other types of transition.

\end{abstract}
\newpage \baselineskip=18pt \setcounter{footnote}{0}

\section{Introduction}

A lot of attention has been focused on the properties of the system
consisting of the intersection of $N_c$ color D$p$--branes and $N_f$
flavour D$q$--branes ($p<q$). In the large $N_c\gg N_f$ limit the
D$p$--branes can be substituted by their corresponding black
$p$--brane supergravity background, while the D$q$--branes are in the
probe limit\cite{Karch:2002sh}.

In addition the D$q$--branes are extended along $q-p$ of the $9-p$
dimensions transverse to the D$p$--brane, and as a result their gauge
degrees of freedom are frozen compared to those of the D$p$--brane.
The dynamics of the $p$--$q$ strings (which transform in the
fundamental of the $SU(N_c)$ low energy gauge theory) and the $q$--$q$
strings  are described by the Dirac--Born--Infeld action of the
D$q$--branes.

Among the issues of interest was the study of the thermodynamic
properties of the
dual\cite{Maldacena:1997re,Witten:1998qj,Gubser:1998bc} Yang--Mills
theory and certain thermal phase transitions in the dynamics of the
fundamental matter. The first study of this nature was for the D3/D7
system and was considered in ref.\cite{Babington:2003vm}, where the
authors considered the near--horizon limit of the non--extremal black
3--brane solution, corresponding to the AdS$_5$-BH$\times S^5$ geometry
(the anti--de Sitter (AdS) spacetime contains a black hole). The
D7--brane wraps a $S^3\subset S^5$ and extends in the radial direction
of the AdS$_5$--BH. The size of the $S^3$ varies as a function of the
radial coordinate.  The D7--brane embeddings then naturally form two
classes: embeddings that reach the horizon and hence fall into the
black hole, and embeddings for which the wrapped $S^3$ shrinks to zero
size at some radial position. For these, the D7--brane world--volume
simply closes smoothly before the horizon. In an Euclidean
presentation, the compact, unbounded parts of the D7--brane have the
topology $S^3\times S^1$ since the Euclidean time has a periodicity
set by the inverse temperature of the system. The classes are then
distinguished by one or the other compact space shrinking away.  The
authors of ref.\cite{Babington:2003vm} proposed that the (topology
changing) transition of the D7--brane embeddings corresponds to a type
of  confinement/deconfinement phase transition, now in the meson
sector of the theory. This system has been extensively studied in
refs.\cite{Apreda:2005yz}--\cite{Mateos:2007vc} and it was shown that
it is a first order phase transition providing a holographic
description of the meson melting phase transition of the fundamental
matter.

There is also a unique critical embedding separating those two
classes. This solution reaches the horizon {\it and} has a shrinking
$S^3$. It has a conical singularity. Solutions of this type will
occupy much of our attention in this paper. Many of these features
generalize to the general D$p$/D$q$ system.  In
ref.\cite{Mateos:2006nu} the D$p$/D$q$ system was considered and some
universal properties, associated with this critical solution
separating the two classes of embedding, were uncovered. In particular
it was shown that for a certain temperature the theory exhibits a
discrete self--similar behavior, manifested by a double logarithmic
spiral in the solution space. This space of solutions is parameterized by
the bare quark mass and the fermionic condensate. (Geometrically
these correspond, respectively, to the asymptotic separation of the
D7-- and D3-- branes and the degree of bending of the D7--branes away
from the D3--branes.)

The region of solution space where the self--similar spiral is located
is unstable, in fact: There is a first order phase transition
associated with the physics of the system jumping between branches of
solutions and bypassing it entirely.  Nevertheless, it seems that
important features of the full physical story can be captured by
examining the neighbourhood of this critical solution.  It is
remarkable that the critical exponents (or better ``scaling
exponents'', so as not to confuse the physics with the nomenclature of
second order phase transitions) characterizing this logarithmic
structure exhibit universal properties and depend only on the
dimension of the internal $S^n$  wrapped by the D$q$--brane.
The precise value of the critical temperature is irrelevant. The
structure is determined by focusing on the local geometry near the
conical singularity of the critical D$q$--brane embedding, and the
exponents are then naturally determined by the study of possible
embeddings in a Rindler space\cite{Frolov:2006tc,Mateos:2006nu}.

The studies described above concern a thermally driven phase
transition. As the temperature passes a certain threshold, thermal
fluctuations seek out the new global minimum that appears and the
system undergoes a transition to a new phase.  In this paper we study
transitions of the system under the effect of two different types of
control parameters: an external electric field and an R--charge
chemical potential, revisiting work done on these systems in
refs.\cite{Albash:2007bq,Albash:2006bs}.  We show that the
corresponding scaling exponents are again universal and depend only on
the dimension of the internal sphere wrapped by the D$q$--brane.  We
find that the key properties of the critical solution can be
determined from the local properties of the geometry, and we find that
this geometry arises naturally by working in a rotating frame, arrived
at using T--duality.  The resulting physics is not controlled by
thermal dynamics, the local geometry is not Rindler, and so the
exponents are different.  The phase transition is driven by the
quantum (as opposed to thermal) fluctuations of the system, as can be
seen from the fact that they persist at zero temperature. It is
satisfying that we can cast these different types of transition into
the same classifying framework.

The structure of the paper is as follows. We begin in section~2 by
reviewing the results of refs.\cite{Frolov:2006tc,Mateos:2006nu}  for
the thermally driven phase transition, focusing on the structure of
the unstable critical solution, extracting the universal properties of
the corresponding scaling exponents. We highlight the natural
appearance of a Rindler geometry.

In section~3.1 we consider the case of an external electric field and
in the {\it insulator/conductor} phase transition discussed
in ref. \cite{Albash:2007bq}. By employing an appropriate T--dual
description of the system we demonstrate that the structure of the
instability and the scaling exponents can be naturally studied by
classifying the possible embedding in a flat {\it rotating} frame. We
observe that these scaling exponents are again universal and depend
only on the dimension of the internal $S^n$ sphere, wrapped by the
D$q$--branes.

In section~3.2 we consider instead the presence of a finite R--charge
chemical potential in the D$p$/D$q$ system and demonstrate that the
resulting phase transition has the same scaling exponents as the
insulator/conductor phase transition driven by an external electric
field.

We consider some generalizations of the discussion in section~4 and
close with some remarks in section~5.

\section{Thermal Phase Transition}

Let us begin by reviewing the result of
refs.\cite{{Frolov:2006tc},{Mateos:2006nu}}. We will be using the
notations of ref.\cite{Mateos:2006nu}. Consider the near--horizon
black D$p$--brane given by:
\begin{eqnarray}\label{1}
&&ds^2=H^{-\frac{1}{2}}\left(-f dt^2+\sum\limits_{i=1}^{p}dx_i^2\right)+H^{\frac{1}{2}}\left(\frac{du^2}{f}+u^2d\Omega_{8-p}^2\right)\ , \\
&&e^{\Phi}=g_s H^{(3-p)/4}\ ,~~~C_{01\dots p}=H^{-1}\ ,\nonumber
\end{eqnarray}
where $H(u)=(R/u)^{7-p}$, $f(u)=1-(u_H/u)^{7-p}$ and $R$ is a length
scale (the AdS radius in the $p=3$ case). According to the
gauge/gravity correspondence, string theory on this background is dual
to a $(p+1)$--dimensional gauge theory at finite temperature. Now if
we introduce D$q$--brane probe having $d$ common space-like directions
with the D$p$--brane, wrapping an internal $S^n\subset S^{8-p}$ and
extended along the holographic coordinate $u$, we will introduce
fundamental matter to the dual gauge theory that propagates along a
$(d+1)$--dimensional defect.

\begin{figure}[h] 
    \centering \includegraphics[width=1.2in]{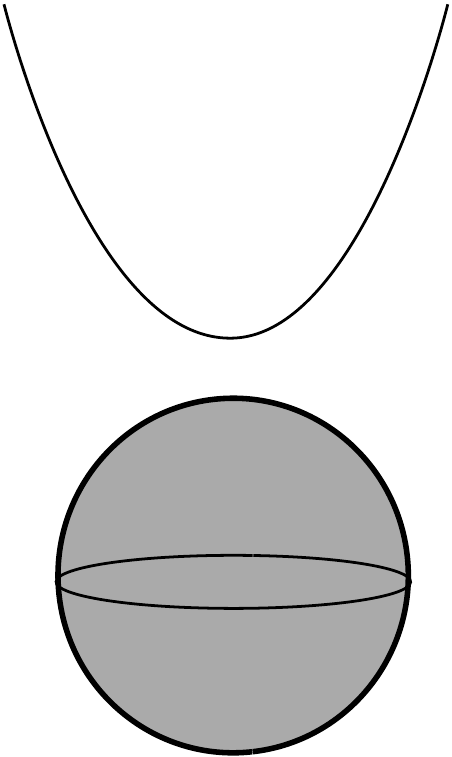}
    \includegraphics[width=1.2in]{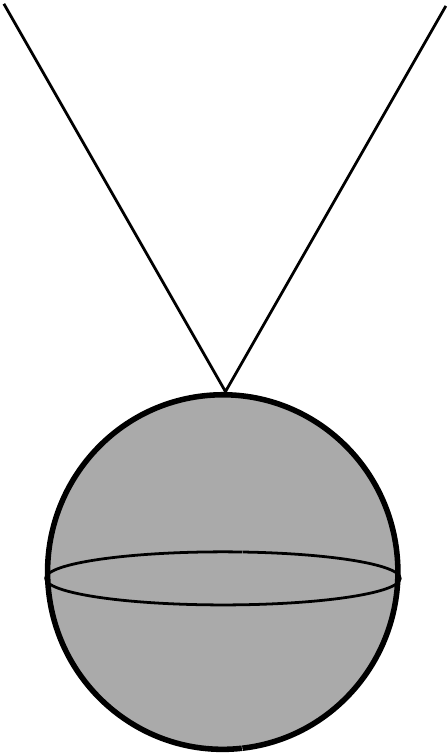}
    \includegraphics[width=1.45in]{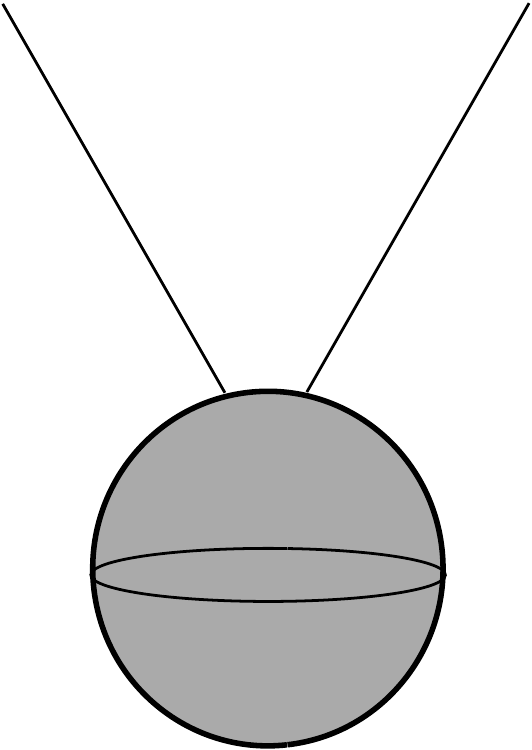}
     \caption{\small \small Schematic diagram depicting the Minkowski (left) and black hole (right) embedding solutions that are separated by a ``critical'' embedding (centre), which has a conical singularity at the event horizon.}
     \label{fig:A1}
  \end{figure}

If we parameterize $S^{8-p}$ by:
\begin{equation}
d\Omega_{8-p}^2=d\theta^2+\sin^2\theta d\Omega_n^2+\cos^2\theta d\Omega_{7-p-n}^2\ ,
\end{equation}
where $d\Omega_m^2$ is the metric on a round unit radius $m$--sphere,
the DBI part of the Lagrangian governing the classical embedding of
the probe is given by\footnote{We consider only systems T-dual to the D3/D7 one, which imposes the constraint $p-d+n+1=4$.}:
\begin{equation}
{\cal L}\propto e^{-\Phi}\sqrt{-|g_{\alpha\beta}|}=\frac{1}{g_s}u^n\sin^n\theta\sqrt{1+f u^2\theta'^2}
\end{equation}
The embeddings split to two classes of different topologies:
``Minkowski'' embeddings, which have a shrinking $S^n$ above the
vanishing locus (the horizon) and yield the physics of meson states
and ``black hole'' embeddings that reach the vanishing locus,
corresponding to a melted/deconfined phase of the fundamental matter.
These classes are separated by a critical embedding with a conical
singularity at the vanishing locus, as depicted in figure
\ref{fig:A1}.

\begin{figure}[h] 
    \centering \includegraphics[width=3.1in]{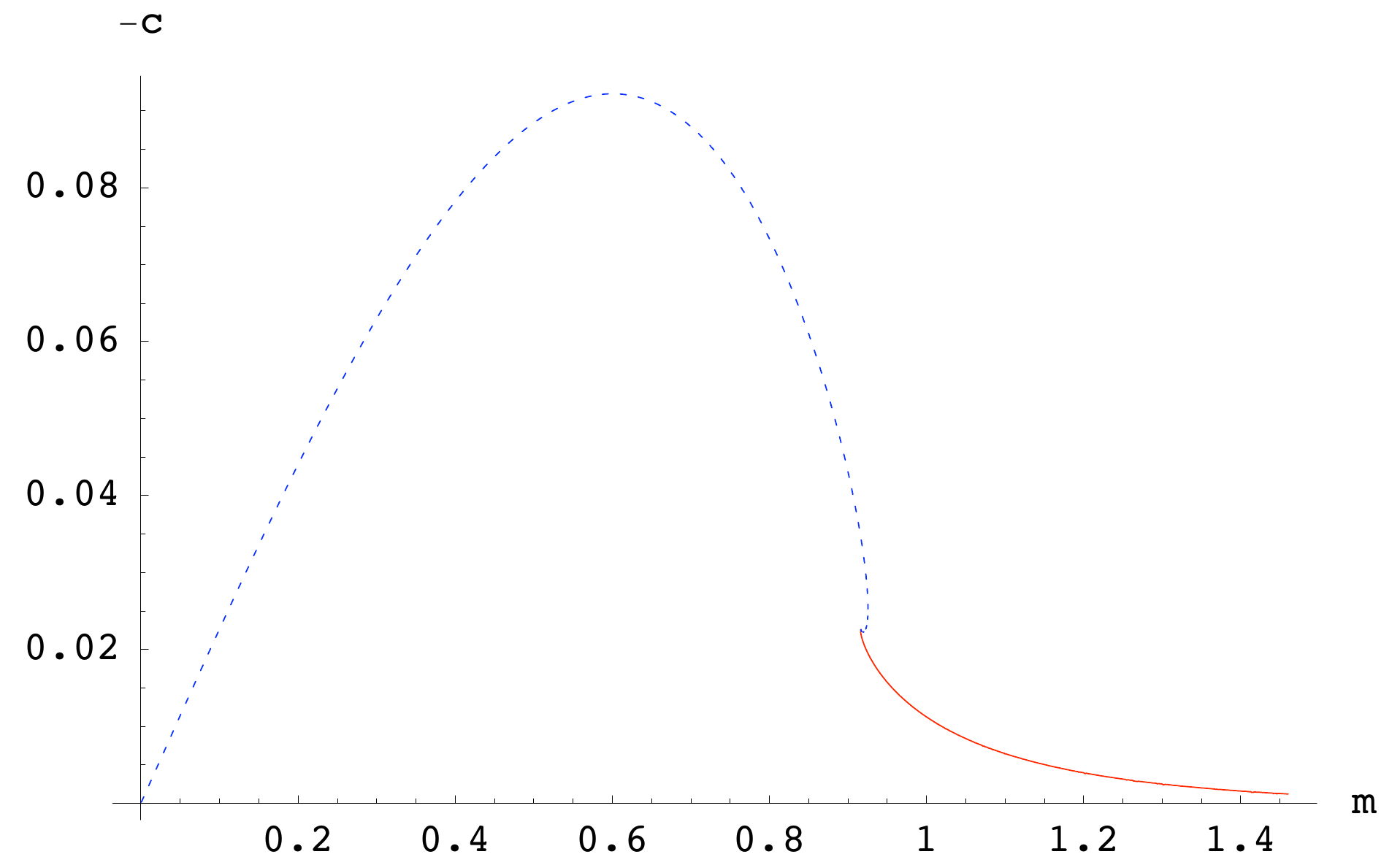}
    \includegraphics[width=3.1in]{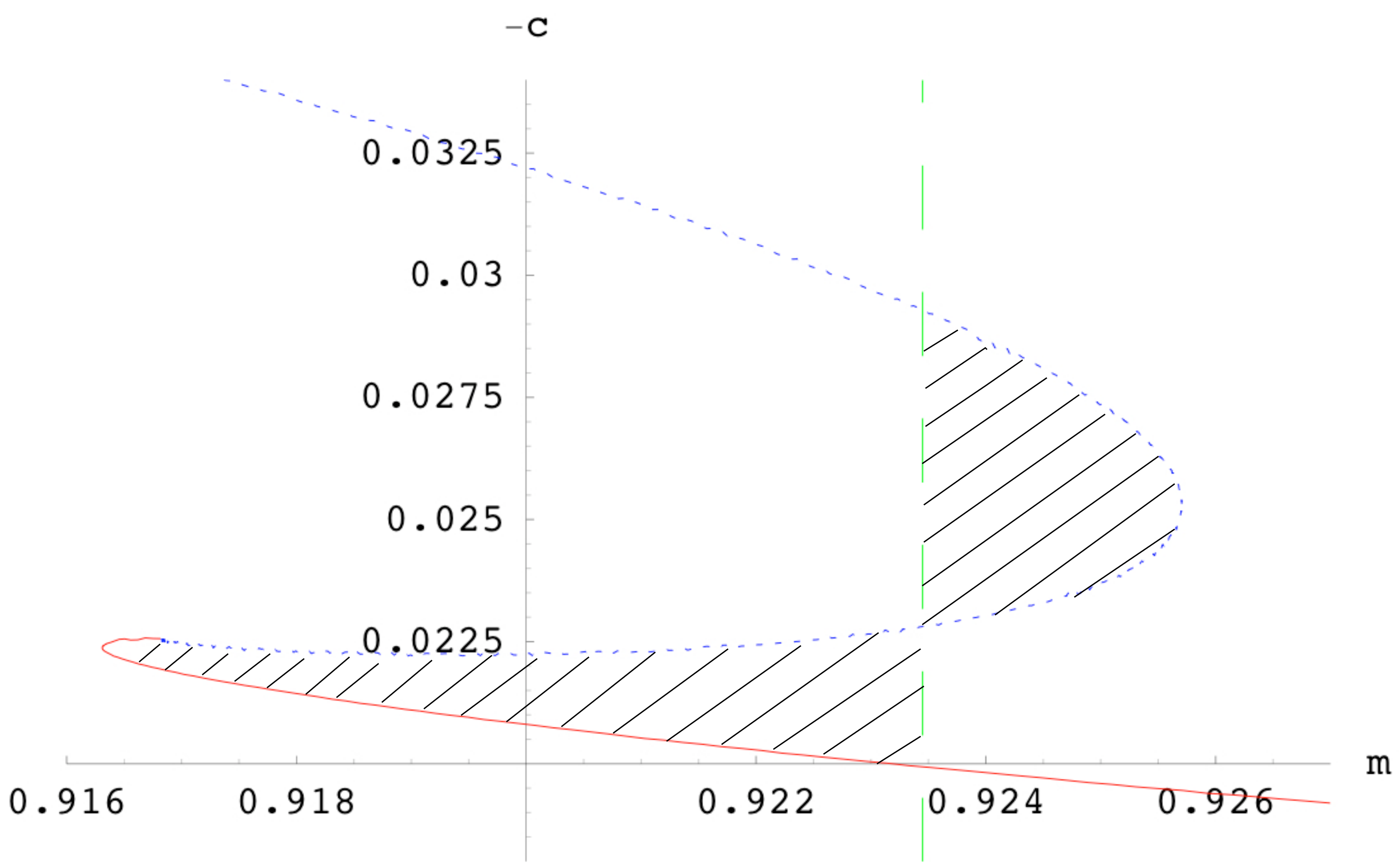}
     \caption{\small \small Plot of the equation of state $c(m)$. The zoomed region shows the location of the first order phase transition. There is a spiral structure hidden near the ``critical'' solution in the neighbourhood of $m=0.9185$, $-c=0.0225$}
     \label{fig:A2}
  \end{figure}

  It is convenient to introduce the following coordinates:
\begin{eqnarray}
&& r^{\frac{7-p}{2}}=\frac{1}{2}\left(u^{\frac{7-p}{2}}+\sqrt{u^{7-p}-u_H^{7-p}}\right)\ ,\\
&& L=r\cos\theta\ ,\quad\mathrm{and}\quad \rho=r\sin\theta\ .  \nonumber
\end{eqnarray}
Then one can show\cite{Karch:2002sh,Kruczenski:2003uq} that the
asymptotic behavior of the embedding at $\rho\to\infty$ encodes the
bare quark mass $m_q={m}/{2\pi\alpha'}$ and the quark condensate
$\langle\bar\psi\psi\rangle\propto-c$ of the dual gauge theory via the
expansion:
\begin{equation}
L(\rho)=m+\frac{c}{\rho^{n-1}}+\dots
\end{equation}
After solving numerically for each embedding of the D$q$--brane, the
parameters $m$ and $c$ can be read off at infinity. From the full
family of embeddings, a plot of the equation of state of the system
$c(m)$ can be generated. The resulting plot for the D3/D7
system\cite{Albash:2006ew} is presented in figure \ref{fig:A2}. The
two different colors (and line types) correspond to the two different
classes of embeddings. The equation of state is a multi--valued
function, and there is a first order phase transition when the free
energies of the uppermost and lowermost branches match.

  The main subject of our discussion is the spiral structure in the
  solution space near the critical
  embedding\cite{Mateos:2006nu,Mateos:2007vn}. In the enlarged portion
  on the right in figure~\ref{fig:A2} it is located to the lower left,
  roughly at $m=0.9185$,  $-c=0.0225$. The spiral structure that is
  hidden near this point is a signal of the discrete self--similarity
  of the theory near the critical solution.

  In order to understand the origins of the spiral, we zoom into the
  space--time region near the tip of the cone of the critical
  embedding\cite{Frolov:2006tc,Mateos:2006nu} using the change of variables:
\begin{equation}
u=u_H+\pi T z^2\ ;~~~\theta=\frac{y}{R}\left(\frac{u_H}{R}\right)^{\frac{3-p}{4}}\ ;~~~\hat x=x\left(\frac{u_H}{R}\right)^{\frac{7-p}{4}}\ .
\end{equation}
Here $T$ is the temperature of the background given by:
\begin{equation}
{T}=\frac{7-p}{4\pi R}\left(\frac{u_H}{R}\right)^{\frac{5-p}{2}}\ .
\end{equation}
 Leaving only the
leading terms in $z$ results in the following metric:
\begin{equation}
ds^2=-(2\pi T)^2z^2 dt^2+dz^2+dy^2+y^2 d\Omega_n^2+d\hat x_d^2+\dots
\label{Rindler}
\end{equation}

The metric (\ref{Rindler}) corresponds to flat space in Rindler
coordinates. The embeddings of the D$q$--branes in the background
(\ref{Rindler}) again split into two different classes: Minkowski
embeddings characterized by shrinking $S^n$ ($y=0$) at some finite
$z_0$, and black hole embeddings, which reach the horizon at $z=0$
for some finite $y=y_0$ (the radius of the induced horizon). The
equation of motion is derived from the Dirac--Born--Infeld action of
the D$q$--branes, which has the following Lagrangian:
\begin{equation}
{\cal L}\propto z y^n\sqrt{1+y'^2}\ .
\end{equation}
The equation of motion derived from this reads:
\begin{equation}
zyy''+(yy'-nz)(1+y'^2)=0\ .
\label{scaling}
\end{equation}
Solutions of this equation enjoy the scaling property $y(z)\to
\frac{1}{\mu}y(\mu z)$, in the sense that if $y(z)$ is a solution to
the equation (\ref{scaling}) so is $\frac{1}{\mu}y(\mu z)$. Under such
a re--scaling the initial conditions $(z_0,y_0)$ for the two classes
of embeddings scale as:
\begin{equation}
z_0\to z_0/\mu;~~~y_0\to y_0/\mu;
\label{re-scaling}
\end{equation}
This suggests the existence of a critical solution characterized by
$z_0=y_0=0$. One can check that $y=\sqrt{n}z$ is the critical
solution. It  has a conical singularity at $y=z=0$.

To analyze the parameter space of the solutions we can linearize the
equation of motion (\ref{scaling}) near the critical solution by
substituting $y(z)=\sqrt{n}z+\xi(z)$, for small $\xi(z)$. The
resulting equation of motion is:
\begin{equation}
z^2\xi''(z)+(n+1)(z\xi'(z)+\xi(z))=0\ ,
\end{equation}
which has a general solution of the form:
\begin{eqnarray}
&&\xi(z)=\frac{1}{z^{r_n}}(A\cos({\alpha_n\ln z)}+B\sin(\alpha_n\ln z))\ ,\label{linsol}\\
&&\mathrm{with}\quad r_n=\frac{n}{2}\ ;~~~\alpha_n=\frac{1}{2}\sqrt{4(n+1)-n^2}\ .\nonumber
\end{eqnarray}
Note that $\alpha_n$ are real only for $n\le4$, which are the cases
naturally realized in string theory\cite{Mateos:2007vn}. Now the
scaling property of equation (\ref{scaling}), combined with the form of
the solutions (\ref{linsol}) suggests the following transformation of
the parameters $(A,B)$ under the re-scaling of the initial conditions
given in equation~(\ref{re-scaling}):
\begin{equation} 
\begin{pmatrix} A' \\  B' \end{pmatrix}=\frac{1}{\mu^{r_n+1}}\begin{pmatrix}\cos{(\alpha_n\ln\mu)} & \sin{(\alpha_n\ln\mu)}\\-\sin{(\alpha_n\ln\mu)} & \cos(\alpha_n\ln\mu)\end{pmatrix}\begin{pmatrix} A \\  B \end{pmatrix}\ .
\label{transformation}
\end{equation}
For a fixed choice of the parameters $A$ and $B$, the parameters $(A',
B')$ describe a double spiral, whose step and periodicity are set by
the real and imaginary parts of the critical/scaling exponents $r_n$
and $\alpha_n$.  

Equation~(\ref{scaling}) has a $Z_2$ symmetry\cite{Frolov:2006tc}
relating the two classes of solutions (Minkowski and black hole
embeddings). If the parameters $(A, B)$ describe
one class of embeddings, then the parameters $(-A,-B)$ describe the
other. In this way the full parameter space near the critical solution
(given by $A=0$, $B=0$) is a double logarithmic spiral.

This self--similar structure of the embeddings near the critical
solution in our Rindler space is transferred by a linear transformation
to the structure of the solutions in the $(m, c)$ parameter space.  If
we call $(m^*, c^*)$ the parameters corresponding to the critical
embedding from figure~\ref{fig:A1}, then sufficiently close to the
critical embedding we can expand:
\begin{equation}
\begin{pmatrix}
m-m^*\\c-c^*
\end{pmatrix}=M\begin{pmatrix}A\\B\end{pmatrix}+O(A^2)+O(B^2)+O(A,B)\ .
\end{equation}
The constant matrix $M$ cannot be determined analytically and depends
on the properties of the system. Generically it should be invertible
(numerically we have verified that it is) and therefore in the
vicinity of the parameter space close to the critical embedding
$(m^*,c^*)$ there is a discrete self--similar structure determined by
the transformation:
\begin{equation}
\begin{pmatrix} m'-m^*\\  c'-c^* \end{pmatrix}=\frac{1}{\mu^{r_n+1}}M\begin{pmatrix}\cos{(\alpha_n\ln\mu)} & \sin{(\alpha_n\ln\mu)}\\-\sin{(\alpha_n\ln\mu)} & \cos(\alpha_n\ln\mu)\end{pmatrix}M^{-1}\begin{pmatrix} m-m^*\\  c-c^* \end{pmatrix}\ . 
\label{fultrans}
\end{equation}
Let us define two solutions to be ``similar'' if:
\begin{equation}
\begin{pmatrix}|m'-m^*|\\|c'-c^*|\end{pmatrix}=\frac{1}{\mu^{r_n+1}}\begin{pmatrix} |m-m^*| \\ |c-c^*|\end{pmatrix}\ .
\end{equation}   
Then one can see from equation (\ref{fultrans}) that this is possible
only for a discrete set of $\mu$s given by:
\begin{equation}
\mu=e^{{k\pi}/{\alpha_n}};~~~k=1, 2, \dots
\end{equation}
Note that in general the matrix $M$ in equation (\ref{fultrans}) will
deform the spiral structure given by the transformation
(\ref{transformation}). However the scaling properties of the theory
remain the same as they are completely determined by the scaling
exponents: $r_n,\alpha_n$. Furthermore one can see that the scaling
exponents depend only on the dimension of the internal sphere $S^n$
wrapped by the D$q$--brane and are thus universal, in the sense that
the detailed value of the critical temperature is irrelevant. It is
the spiral structure that ultimately seeds the multi--valuedness of the
space of solutions, twisting the $(m,-c)$ curve back on itself as in
figure~\ref{fig:A2}. Therefore, it is the spiral --- and the
neighbourhood of the critical solution from where it emanates --- that
is responsible for the presence of a first order phase transition in
the system. Whether there is a spiral or not can be read off from the
scaling parameters $(r_n,\alpha_n)$, and since\cite{Mateos:2007vn} for
all consistent D$p$/D$q$ systems the condition $n\le 4$ is satisfied
the corresponding thermal phase transition (meson melting at large
$N_c$) is a first order one.

\section{Quantum--Induced  Phase Transitions}

In this section we will consider a different class of phase
transitions. These are arise in the presence of external fields, and
can happen even at zero temperature, and so since the fluctuations
driving the transition are no longer thermal, they might be expected
to be in a different class.  Naively, the broad features of the
equation of state --- multi--valuedness and so forth --- have
similarities with the thermal case, and so it is natural to attempt to
trace the extent to which these similarities persist. We will find
that once we cast these systems in the language of the previous
section, the similarities and differences will be quite clear.

We will first concentrate on the case of an external electric field.
The flavoured system, at large enough electric field, has an {\it
  insulator/conductor} phase transition, as studied in
ref.\cite{Albash:2007bq}. As with the thermal transition of the last
section, the mesons dissolve into their constituent quarks, but this
time it is due to the electric field overcoming their binding energy.
The transition is of first order.

As we saw in the previous section the scaling properties of the
thermally driven phase transition are naturally studied in a Rindler
frame with a temperature set by the temperature of the background.  In
ref.\cite{Albash:2007bq} it was shown that in analogy to the thermally
driven phase transition there is a nice geometrical description of the
electrically driven phase transition, and the structure of the system
can be again characterized by an unstable critical embedding with a
conical singularity at an appropriate vanishing locus (analogous to
the event horizon). Here, we will generalize this description to the
case of the D$p$/D$q$ system.

Furthermore after an appropriate T-duality transformation we will show
that the vanishing locus corresponds to an effective ``ergosphere''
due to a rotation of the coordinate frame along the compact directions
of the background. The instability near criticality is then naturally
interpreted as an instability due to the over--spinning of the
D$(q-1)$ brane probes (in the T--dual background) as they reach the
ergosphere. We then study the structure of the theory near criticality
by zooming in on the space--time region in the vicinity of the conical
singularity. Once again, we will find that the structure is entirely
controlled by the dimension of the internal sphere, $S^n$,  wrapped by
the D$(q-1)$--branes (in the T--dual background) --- details such as
the value of the electric field and the temperature of the system, are
irrelevant.

\subsection{Criticality and Scaling in an External Electric Field.}
Let us consider the near--horizon black D$p$--brane given by the
background in equation~(\ref{1}). Following a similar
idea\cite{Filev:2007gb} for producing a background magnetic field, if
we turn on a pure gauge $B$--field in the $(t,x_p)$
plane\cite{Karch:2007pd,Albash:2007bq,Erdmenger:2007bn}, in the dual
gauge theory this will correspond to an external electric field,
oriented along the $x_p$ direction:
\begin{equation}
B=E dt\wedge dx_p\ .
\label{beefield}
\end{equation}
The resulting Lagrangian is:
\begin{equation}
{\cal L}\propto e^{-\Phi}\sqrt{-|g_{\alpha\beta}+B_{\alpha\beta}|}=\frac{1}{g_s}\sqrt{\frac{f-E^2H}{f}}u^n\sin^n\theta\sqrt{1+f u^2\theta'^2}\ .
\label{electric}
\end{equation}
This leads to the existence of a vanishing locus at $u=u_*$ given by:
\begin{equation}
u_*^{7-p}=u_H^{7-p}+E^2L^{7-p}\ ,
\label{locus}
\end{equation}
at which the action (\ref{electric}) vanishes. Notice that this is
distinct from the horizon, and even at zero temperature will be
present. A study of the local physics near this locus will therefore
pertain to non--thermal physics.

The embeddings split into two different classes: Minkowski
embeddings which have a shrinking $S^n$ above the vanishing locus and
correspond to meson states and embeddings reaching the vanishing
locus, corresponding to a deconfined phase of the fundamental matter.
These classes are separated by a critical embedding with a conical
singularity at the vanishing locus. Our goal is to explore the
self--similar behavior of the theory near this critical embedding and
calculate the corresponding scaling exponents.

In order to make the analysis closer to the one performed in
refs.\cite{{Frolov:2006tc},{Mateos:2006nu}}, for the thermal phase
transition (described in the last section), we T--dualize along the
$x_p$ direction.  This is equivalent to a trading of the pure gauge
$B$--field for a rotating frame in the T--dual background. Indeed the
geometry T--dual to equation~(\ref{1}), with the $B$--field given by
equation~(\ref{beefield}), is given by:
\begin{eqnarray}
&&d\tilde s^2=H^{-\frac{1}{2}}(-\tilde f dt^2+\sum\limits_{i=1}^{p-1}dx_i^2)+2H^{\frac{1}{2}}E dt d{\tilde x}_p+H^{\frac{1}{2}}\left(\frac{du^2}{f}+u^2d\Omega_{8-p}^2+d{\tilde x}_p^2\right)\ ,\label{tdual}\\
&&e^{\tilde\Phi}=g_sH^{1-\frac{p}{4}};~~~\tilde f=1-\left(\frac{u_*}{u}\right)^{7-p}\ .\nonumber
\end{eqnarray}
The background given by equation (\ref{tdual}) corresponds to the
near--horizon limit of a stack of $N_c$ D$(p-1)$--branes smeared along
the coordinate $\tilde x_p$. Now if we place a probe D$(q-1)$--brane
having $(d{-}1)$ spatial directions shared with the D$(p-1)$---branes,
filling the radial direction $u$ and wrapping an internal $S^n$ inside
the $S^{8-p}$ sphere of the background, we will recover the action
(\ref{electric}), as we should.

Note that in these coordinates we have an effective ``ergosphere''
coinciding with the vanishing locus given by equation~(\ref{locus}).
Now the critical embedding is the one touching the ergosphere and
having a conical singularity at $u=u_*$. In the $(m, c)$--plane this
embedding corresponds to the center of the spiral structure $(m_*,
c_*)$.

Despite the analogy with the analysis of the thermal phase transition,
in this case there is a crucial difference, because of the necessity
(from charge conservation) for the D$(q-1)$--brane to extend beyond
the ergosphere. Indeed since the D$(q-1)$--brane is an extended object
one can find static solutions that extend beyond the ergosphere and
are non--superluminal. To this end one should allow the
D$(q-1)$--brane to extend along the direction of rotation $\tilde
x_p$.  In the original coordinates (before T--dualization) this is
equivalent to a non--trivial profile for the $A_p$ component of the
gauge field, which corresponds to the appearance of a global electric
current along the
$x_p$--direction\cite{{Karch:2007pd},{Albash:2007bq}}. This is the
reason why we refer to the corresponding phase transition as an
insulator/conductor phase transition. After the transition, the quarks
are free to flow under the influence of the electric field, forming a
current.

Let us describe how this procedure works in the case of a general
D$(p-1)$/D$(q-1)$--intersection. Again we will work in the T-dual
background (\ref{tdual}).  Let us consider an ansatz for the
D$(q-1)$--brane embedding of the form:
\begin{equation}
\theta=\theta(u)\ ;~~~{\tilde x}_p={\tilde x}_p(u)\ ;
\label{ext}
\end{equation}
this leads to the action:
\begin{equation}
{\cal L}_*\propto\frac{1}{g_s}\sqrt{\frac{f-E^2H}{f}}u^n\sin^n\theta\sqrt{1+f u^2\theta'^2+\frac{f^2}{f-E^2H}{{\tilde x}'^2_p}}\ .
\end{equation}
Now after integrating the equation of motion for $\tilde x_p$ and
plugging the result in the original Lagrangian, we get the following
on--shell Lagrangian:
\begin{equation} 
{\cal L}_*\propto\frac{1}{g_s}\sqrt{\frac{f-E^2H}{f u^{2n}\sin^{2n}\theta-K^2}}u^{2n}\sin^{2n}\theta\sqrt{1+f u^2\theta'^2}\ .
\label{action*}
\end{equation}
It is easy to verify that if we choose the integration constant $K^2$
in equation~(\ref{action*}) to satisfy:
\begin{equation}
K^2=E^2H_{*}u_*^{2n}\sin^{2n}\theta_0\ ,
\label{inconst}
\end{equation}
then the action (\ref{action*}) is regular at the ergosphere
($u=u_*$). Note that at the critical embedding
$\theta_0=\theta_*\equiv0$ and the constant in
equation~(\ref{inconst}) is zero. This constant is proportional to the
global electric current along the $x_p$ direction of the original
D$p$/D$q$--brane system. (See
refs.\cite{{Karch:2007pd},{Albash:2007bq}} for a discussion in the
case of the D3/D7 system.)

We are interested in the scaling properties of the theory, near the
critical embedding solution. Despite the fact that the Lagrangians
(\ref{electric}) and (\ref{action*}) describing the Minkowski and
ergosphere classes of embeddings are different, the fact that at
the critical embedding they coincide ($K^2$=0) shows that the
corresponding equations of motion share the same critical solution.
Furthermore, as we will see, the critical exponents are the
same for both types of embedding. 

Let us introduce dimensionless coordinates by the transformation:
\begin{equation}  
u=u_*+z\frac{Du_*}{7-p}\ ;~~\theta=\frac{y}{R}\left(\frac{u_*}{R}\right)^{\frac{3-p}{4}}\ ;~~x_i\left(\frac{u_*}{R}\right)^{\frac{7-p}{4}}\to x_i\ ;~~t\left(\frac{u_*}{R}\right)^{\frac{7-p}{4}}\to t\ ;~~H_*^{\frac{3}{4}}E\tilde x_p\to\tilde x_p\ ,
\label{zooming}
\end{equation}
where $D^2=(7-p)^2f_*/H_*^{\frac{1}{2}}u_*^2$, $H_*=H|_{u=u_*}$ and
$f_*=f|_{u=u_*}$. To leading order in $z$ and $y$ the metric
(\ref{tdual}) is given by:
\begin{eqnarray}
\label{rotmet}
d\tilde s^2=-Dz dt^2+dz^2+dy^2+y^2d\Omega_n^2+H_*^{\frac{1}{2}}u_*^2d\Omega_{7-p-n}^2+2dtd{\tilde x}_p+\frac{1}{E^2H_*}d{\tilde x}_p^2+\sum\limits_{i=1}^{p-1}dx_i^2\ .
\end{eqnarray}

First consider the case of Minkowski embeddings, characterized by a
distance $z_0$ above the ergosphere at which they close
($y=y(z_0)=0$). The Lagrangian describing the D$(q-1)$--brane
embedding is:
\begin{equation}
{\tilde{\cal L}_*}\propto y^n z^{1/2}\sqrt{1+y'^2}\ ,
\label{lagrscal1}
\end{equation}
The corresponding equation of motion is given by:
\begin{equation}
\partial_z\left(y^n z^{1/2}\frac{y'}{\sqrt{1+y'^2}}\right)-ny^{n-1}z^{1/2}\sqrt{1+y'^2}=0\ .
\label{eqmn}
\end{equation}
Equation (\ref{eqmn}) possesses the scaling symmetry:
\begin{equation}
y\to y/\mu\ ;~~~z\to z/\mu;
\label{scal}
\end{equation}
in the sense that if $y=y(z)$ is a solution to equation (\ref{eqmn})
so is the function $\frac{1}{\mu}y(\mu z)$. Now under the scaling
(\ref{scal}) the boundary condition for the Minkowski embedding
scales as $z_0\to z_0/\mu$. This suggests the existence of a limiting
critical embedding with $z_0=0$, and indeed:
\begin{equation}
y(z)=\sqrt{2n}z\ ,
\label{crit2}
\end{equation}
is a solution to the equation of motion in equation~(\ref{eqmn}).  The
corresponding D$(q-1)$--brane has a conical singularity at $y=z=0$.
Now before we linearize equation (\ref{eqmn}) and calculate the
critical exponents let us consider the case of the ergosphere class of
solutions characterized by the radius of the ergosphere induced on
their world--volume. Because of the possibility to extend beyond the
ergosphere we should consider the analog of the ansatz from
equation~(\ref{ext}):
\begin{equation}
y=y(z);~~~\tilde x_p=\tilde x_p(z)\ .
 \end{equation}
 The corresponding Lagrangian is:
 \begin{equation}
\tilde{\cal L_*}\propto y^n\sqrt{Dz(1+y'^2)+(Fz+1)\tilde x'^2_p}\ ,
\label{ergo-action}
 \end{equation}
 where $F=D/E^2H_*$. After integrating the equation of motion for
 $\tilde x_p$ and substituting it into the Lagrangian
 (\ref{ergo-action}), we obtain the following on--shell Lagrangian:
 \begin{equation}
\tilde{\cal L_*}\propto \frac{z^{1/2}\sqrt{Fz+1}y^{2n}}{\sqrt{(Fz+1)y^{2n}-y_0^{2n}}}\sqrt{1+y'^2}\ .
\label{on-shell-erg}
 \end{equation}
 It is easy to see that the Lagrangian (\ref{on-shell-erg}) is regular
 at $z=0, y=y_0$. The equation of motion for $y(z)$, derived from the
 Lagrangian (\ref{ergo-action}) and with the substituted solution for
 $\tilde x_p(z)$ is:
\begin{equation}
\frac{\partial}{\partial z}\left(\frac{z^{1/2}y'}{\sqrt{1+y'^2}}\sqrt{\frac{(Fz+1)y^{2n}-y_0^{2n}}{Fz+1}}\right)-ny^{2n-1}z^{1/2}\sqrt{\frac{Fz+1}{(Fz+1)y^{2n}-y_0^{2n}}}\sqrt{1+y'^2}=0\ .
\label{eom-erg}
\end{equation}
It is easy to check that equation (\ref{crit2}) is a solution to
equation (\ref{eom-erg}). Furthermore for $z\ll 1/F$ one can see that
equation (\ref{eom-erg}) has the scaling symmetry (\ref{scal}) (note
that equation (\ref{scal}) suggests $y_0\to y_0/\mu$). Linearizing
equations (\ref{eqmn}) and (\ref{eom-erg}) near the critical solution
(\ref{crit2}) by substituting:
\begin{equation}
y(z)=\sqrt{2n}z+\xi(z)
\end{equation}
results in the same equation:
\begin{equation}
z^2\xi''(z)+(n+1/2)(z\xi'(z)+\xi(z))=0\ .
\label{lin2}
\end{equation}
The general solution of equation (\ref{lin2}) is given by:
\begin{equation}
\xi(z)=\frac{1}{z^{r_n}}(A\cos(\alpha_n\ln z)+B\sin(\alpha_n\ln z))\ ,
\end{equation}
where the scaling exponents are given by:
\begin{equation}
r_n=\frac{2n-1}{4}\ ;~~~\alpha_n=\frac{1}{4}\sqrt{7+20n-4n^2}\ .
\label{crit-exp2}
\end{equation}
Note that the scaling exponents again, while quite different from
those of the thermal case (see equation~(\ref{linsol})) depend only on
the dimension of the internal $S^n$ wrapped by the D$q$--brane and are
thus universal for all D$p$/D$q$ systems. Furthermore the discrete
self--similarity holds for $n\leq5$. By similar reasoning to the
thermal case\cite{Mateos:2007vn}, since for all consistent systems
realized in string theory we have that $n\leq4$, for such systems we
may expect that the electrically driven confinement/deconfinement
phase transition is first order and has the described discrete
self--similar behavior near the solution that seeds the
multi--valuedness of the equation of state.

The rest of the analysis is completely analogous to the thermal case
considered in the previous section. Therefore we come to the
conclusion that close to the critical embedding (specified by~$m_*$
and $c_*$) the theory has the following scaling property:
\begin{equation}
\begin{pmatrix} m'-m^*\\  c'-c^* \end{pmatrix}=\frac{1}{\mu^{r_n+1}}M\begin{pmatrix}\cos{(\alpha_n\ln\mu)} & \sin{(\alpha_n\ln\mu)}\\-\sin{(\alpha_n\ln\mu)} & \cos(\alpha_n\ln\mu)\end{pmatrix}M^{-1}\begin{pmatrix} m-m^*\\  c-c^* \end{pmatrix}\ , 
\label{fultrans2}
\end{equation}
with $r_n$ and $\alpha_n$ given by equation (\ref{crit-exp2}).

It is interesting to compare the analytic results some numerical
studies. Let us consider the D3/D7 system. From equation~(\ref{scal})
on can see that the variation of the scaling parameter $\mu$ in
equation~(\ref{fultrans2}) can be traded for the variation of the
boundary conditions of the probe, namely $z_0$ for Minkowski and $y_0$
for ergosphere embeddings. On the other hand, close to the critical
embedding, the change of coordinates in equation (\ref{zooming})
suggests that:
\begin{equation}
\theta_0\propto y_0\quad{\rm and}\quad u_0-u_*\propto z_0\ ,
\end{equation}
where $u_0$ and $\theta_0$ are the boundary conditions for the
embeddings in the original (not zoomed in)
background. Note that the parameter $u_0$ is related to the
constituent quark mass $M_c$ \cite{Mateos:2007vn} (in the absence of
an electric field) via $M_c=(u_0-u_H)/(2\pi\alpha')$. 

Close to the
critical embedding we have that:
\begin{equation}
\mu=(u_{0,{\rm in}}-u_*)/(u_0-u_*)\quad{\rm and}\quad \mu=z_{0,{\rm in}}/z_0\ ,
\end{equation}
for some fixed boundary conditions $u_{0,{\rm in}}$ and
$\theta_{0,{\rm in}}$. Now equation (\ref{fultrans2}) suggests that
for Minkowski embeddings the plot of $(m-m_*)/(u_0-u_*)^{r_n+1}$
versus $\alpha_n\ln(u_0-u_*)$ should be an harmonic function of
$\alpha_n\ln(u_0-u_*)$ with a period $2\pi$. Similarly for ergosphere
embeddings the plot of $(m-m_*)/\theta_0^{r_n+1}$ versus
$\alpha_n\ln\theta_0$ should be a harmonic function of
$\alpha_n\ln\theta_0$ with a period $2\pi$. Note that the physical
meaning of $\theta_0$ can be related to the value of the global
electric current (see equation~(\ref{inconst}) and the comment below).
 
\begin{figure}[h] 
   \centering
      \includegraphics[width=3.2in]{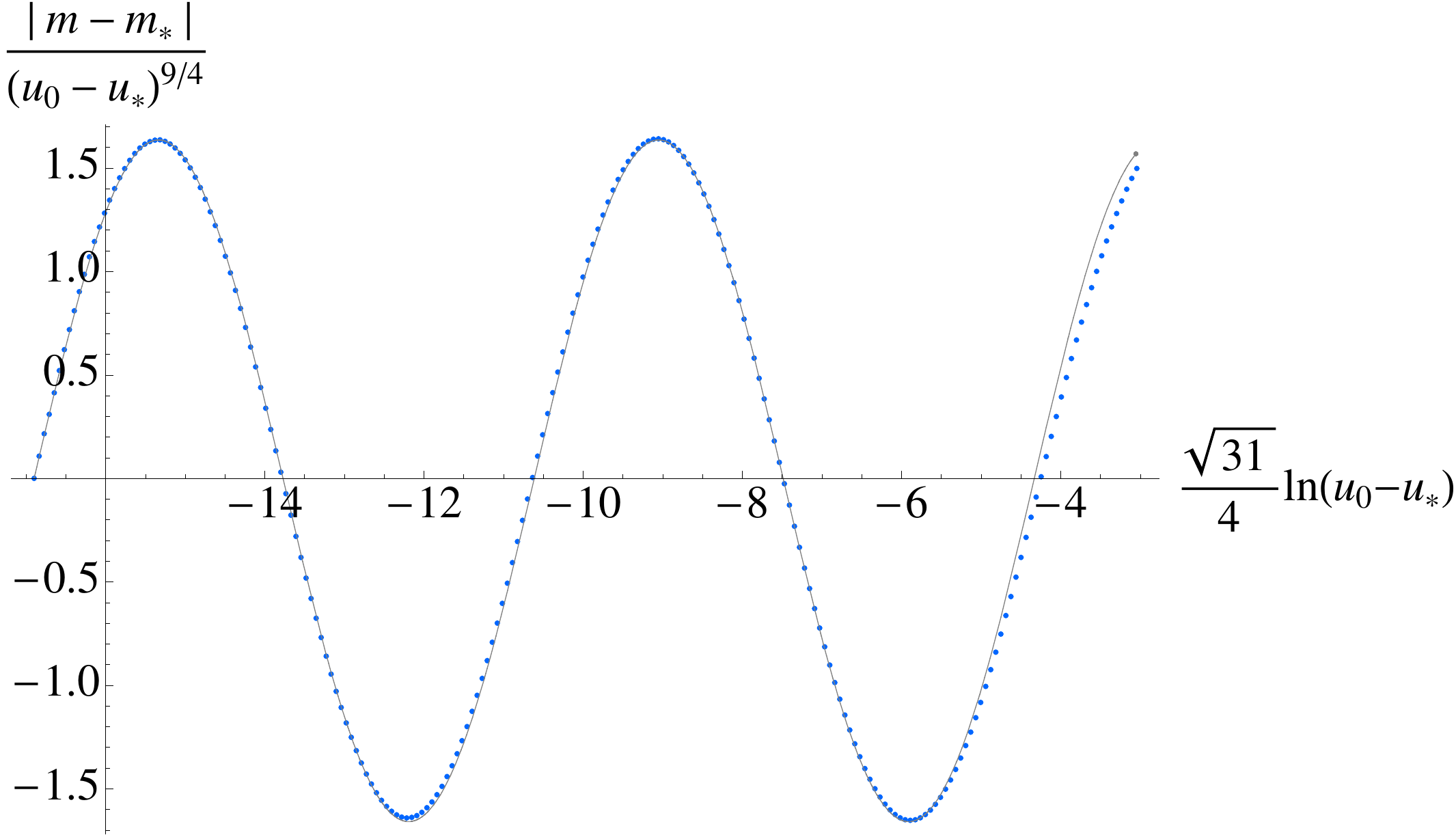} \hskip0.5cm
      \includegraphics[width=3.2in]{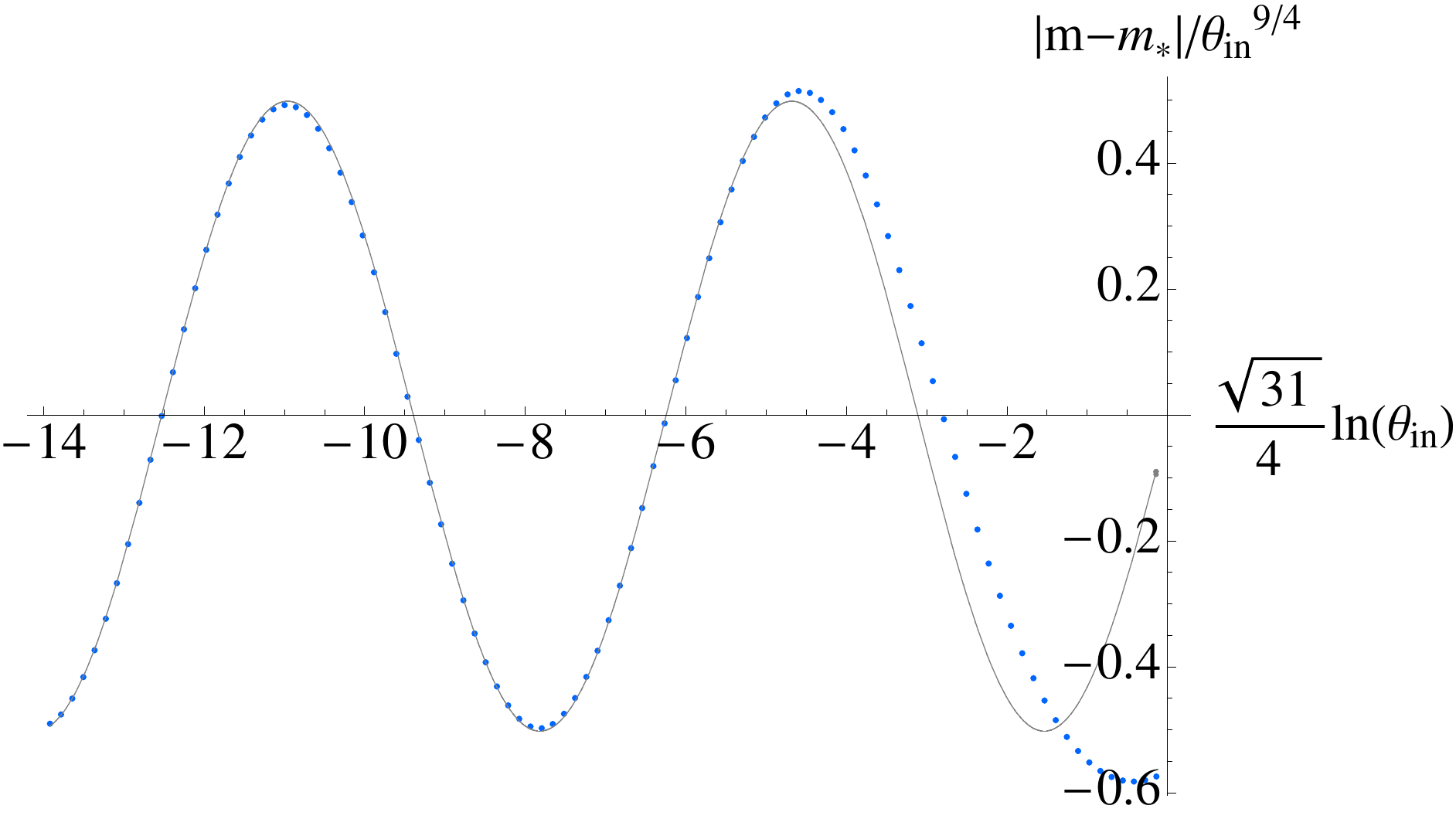} 
   \caption{\small The solid line is a fit with trigonometric functions of period $2\pi$. The plots confirm 
     that the critical exponents of the theory are $r_3=5/4$ and
     $\alpha_3=\sqrt{31}/4$.}
   \label{fig:electric}
\end{figure}

As can be seen in figure \ref{fig:electric}, for both types of
embeddings the numerical results are in accord with
equation~(\ref{fultrans2}) and the analytic results improve deeper
into the spiral (large negative values on the horizontal axis).  Our
numerical results confirm that the critical exponents are indeed
$r_3=5/4$ and $\alpha_3=\sqrt{31}/4$, as the general analytic
results yield.

\subsection{Criticality and Scaling with R--Charge Chemical Potential.}

Now we study the case when the external parameter is an R--charge
chemical potential in the dual gauge theory. We will consider the
system discussed in ref.~\cite{Albash:2006bs}, where a D7--brane probe
in the spinning D3--brane geometry~\cite{Kraus:1998hv,Cvetic:1999xp}
was considered.

The relevant geometry is given by:
\begin{eqnarray}
&&ds^2=\Delta^{1/2}\left(-({\cal H}_1{\cal H}_2{\cal H}_3)^{-1}f dt^2+\frac{u^2}{R^2}d\vec x^2+f^{-1}du^2\right)+\label{spinning}\\
&&\hskip3.0cm +\Delta^{-1/2}\sum_{i=1}^{3}{\cal H}_i\left(\mu_i^2(Rd\phi_i-A_t^idt)^2+R^2 d\mu_i^2\right)\ ,\nonumber
\end{eqnarray}
where
\begin{eqnarray}
&&f=\frac{u^2}{R^2}{\cal H}_1{\cal H}_2{\cal H}_3-\frac{u_H^4}{u^2R^2},~~~{\cal H}_i=1+\frac{q_i^2}{u^2},~~~A_t^i=\frac{u_H^2}{R}\frac{q_i}{u^2+q_i^2},~~~\Delta={\cal H}_1{\cal H}_2{\cal H}_3\sum_{i=1}^{3}\frac{\mu_i^2}{{\cal H}_i},\label{fields}\nonumber\\
&&{\rm with}\qquad\mu_1=\sin\theta,~~~\mu_2=\cos\theta\sin\psi,~~~\mu_3=\cos\theta\cos\psi .
\end{eqnarray}
Here the parameter $u_H$ would be the radius of the event horizon if
the angular momentum of the geometry was set to zero ($q_i=0$). The
radius $u_E$ of the actual event horizon is determined by the largest
root of $f(u)=0$.  The temperature of the background is given
by\cite{Russo:1998by}:
\begin{equation}
T=\frac{u_E}{2\pi R^2 u_H^2}\left(2u_E^2+q_1^2+q_2^2+q_3^2-\frac{q_1^2q_2^2q_3^2}{u_E^4}\right)=\frac{1}{2\pi R^2 u_H^2 u_{E}^{\phantom{2}}}(u_{E}^2-u_1^2)(u_{E}^2-u_2^2)\ ,
\label{temperatureR}
\end{equation}
where $u_1$ and $u_2$ are the other two roots of $f(u)=0$.

The background (\ref{spinning}) has an ergosphere determined by the
expression:
\begin{equation}
\Delta({\cal H}_1{\cal H}_2{\cal H}_3)^{-1}f-\sum_{i=1}^3{\cal H}_i\mu_i^2(A_t^i)^2=0\ .
\label{ergosphere}
\end{equation}
Since the background (\ref{spinning}) is asymptotically AdS$_5\times
S^5$, we can ``remove'' the ergosphere (\ref{ergosphere}), by going to a
rotating frame. This is equivalent to gauge shifting $A_t^i$
from~(\ref{fields}) such that ${A'_t}^i=-\mu_R^i+A_t^i$. The
parameters $\mu_R^i$ are set by the condition ${A'_t}^i|_{u_E}=0$ and
hence:
\begin{equation} 
\mu_R^i=\frac{u_H^2}{R}\frac{q_i}{u_E^2+q_i^2}\ .
\label{chempot}
\end{equation}
From the behaviour at infinity ($u\to\infty$), it is clear that
$\mu_R^i$ correspond to the angular velocities of the frame along
$\phi_i$. In the dual gauge theory these correspond to having time
dependent phases of the adjoint complex scalars or equivalently to
R--charge chemical potentials for the corresponding
scalars\cite{Chamblin:1999tk}.

In order to restore some of the symmetry of the metric
(\ref{spinning}), we will consider the case when $q_2=q_3$. This
corresponds to having an $S^3$ (parameterized by $\psi,\phi_2,\phi_3$)
inside the deformed~$S^5$. Now if we introduce D7--branes filling the
AdS--like part of the geometry and wrapping the~$S^3$, we will add
fundamental matter to the gauge theory. Furthermore we are free to
rotate the D7--branes along~$\phi_1$ and the corresponding angular
velocity is interpreted as a time dependent phase of the bare quark
mass\footnote{We would like to thank A. Karch for pointing this out to
  us.}. (Recall that in introducing D7--branes to the D3--brane system
we actually add flavours as chiral superfields into the ${\cal N}=2$
gauge theory).  If that phase is the same as the phase of the complex
adjoint scalar, $\mu_R^1 t$, it is equivalent to a R--charge
chemical potential for both the adjoint scalar and the chiral field.

On the gravity side of the description this is equivalent to letting
the D7--branes have the same angular velocity $\mu_R^1$ as the
rotating frame of the background. Moving to the frame co--rotating
with the D7--brane corresponds to moving back to the gauge choice for
$A_t^1$ from equation~(\ref{fields}).
The price that we pay is that we again have an
ergosphere this time given by:
\begin{equation}
\Delta({\cal H}_1{\cal H}_2^2)^{-1}f-{\cal H}_1\sin^2\theta(A_t^1)^2=0
\label{ergosphere-new}. 
\end{equation}

The possible D7--brane embeddings then naturally split into two
classes: Minkowski embeddings that have a shrinking $S^3$ above the
ergosphere and ergosphere embeddings which reach the ergosphere. These
classes are again separated by a critical embedding which has a
conical singularity at the ergosphere. In analogy to the T--dual
description of the previous subsection for the external electric field
case, the ergosphere embeddings will have to be extended along
$\phi_1$ so that they can stay non--space--like beyond the
ergosphere\footnote{Note that in \cite{Albash:2006bs} the ergosphere
  class of embeddings are not extended along $\phi_1$.}.  However in
this paper we are interested in the scaling properties of the theory
for parameters $(m, c)$ in the vicinity of the critical parameters
$(m_*,c_*)$, corresponding to the critical embedding. As we saw in the
previous section, modifying the ergosphere class of embeddings so as
to be regular at the ergosphere does not alter the properties of the
theory near the critical solution. In particular the scaling exponents
characterizing the discrete self--similar behavior of the theory
remain the same. So henceforth we will focus on the study of the
Minkowski type of embeddings. The analysis is completely analogous to
the one performed in the previous subsection.

In order to focus on the space--time region close to the conical
singularity of the critical embedding, we consider the change of
coordinates:
\begin{equation}   
u=u_{\rm erg}+\frac{u_Hq_1}{R u_{\rm erg}}z;~~~\cos\theta=\frac{\pi}{2}-\frac{y}{R}\ ,
\end{equation}
where 
\begin{equation}
u_{\rm erg}^2=u_H^2-q_2^2
\end{equation}
is the radial coordinate $u$ of tip of the critical embedding or
equivalently the $\theta=\pi/2$ point of the ergosphere. It can be
shown that for the values of $q_2$ for which the geometry is
not over spun (and so has an horizon) the corresponding value of
$u_{\rm erg}$ is real.

\begin{figure}[h] 
   \centering
      \includegraphics[width=3.2in]{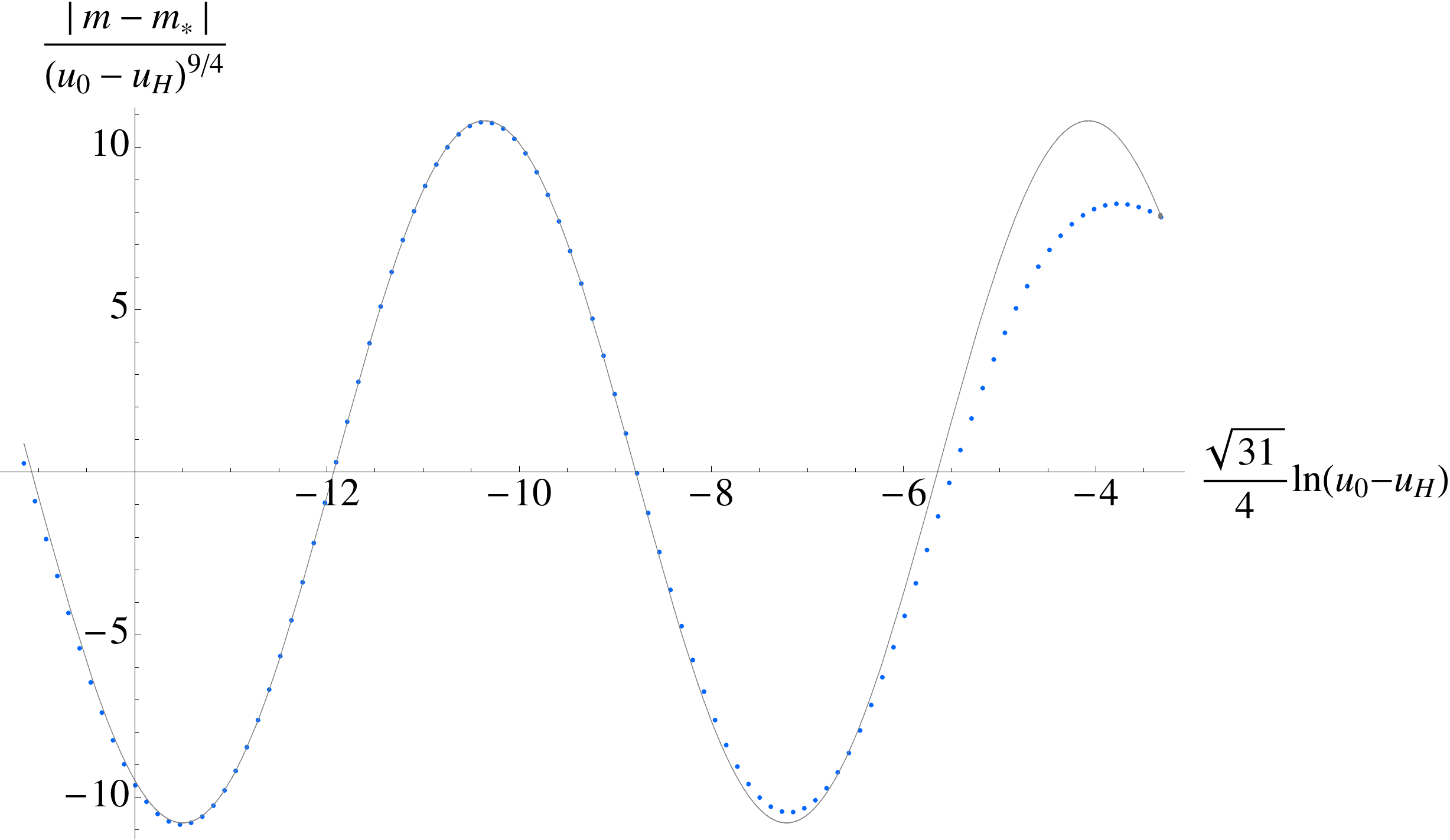} 
    \caption{\small Plot of the relation between the bare quark mass parameter $m$ and the distance above the ergosphere $(u_0-u_H)$. The plot is for $q=0.5$ in units in which $u_H=1$. The solid line is a fit with trigonometric functions of a period $2\pi$. The plot confirms 
      that the scaling exponents of the theory are $r_3=5/4$ and
      $\alpha_3=\sqrt{31}/4$.}
   \label{fig:rcharge}
\end{figure}

After leaving only the leading terms in $z$ and $y$, we get:
\begin{eqnarray}
&&ds^2/\alpha'=-D_1zdt^2+dz^2+dy^2+y^2d\Omega_3^2-2q_1 dt d\phi_1+\frac{u_H^2}{R^2}d\vec x^2+R_1d\phi_1^2\ ,\label{rotmet2}\\
&&d\Omega^3=d\psi^2+\sin^2\psi d\phi_2^2+\cos^2\psi d\phi_3^2\ ;~~~
D_1=\frac{4q_1u_H}{R^3}\ ;~~~R_1^2=\frac{u_{\rm erg}^2+q_1^2}{u_H^2}R^2\ .\nonumber
\end{eqnarray}  
The metric in equation (\ref{rotmet2}) is of the same type as that in
equation~(\ref{rotmet}), namely flat space with some compact
directions in a rotating frame. Therefore the analysis is completely
analogous to the one for the electric case and hence the scaling
exponents are again given by equation (\ref{crit-exp2}) with $n=3$,
because the D7--branes are wrapping an internal $S^3$:
\begin{equation}
r_3=5/4\ ; ~~~\alpha_n=\sqrt{31}/4\ .
\end{equation}
We can again verify this numerically. It is convenient to do this for
the single charge case, namely $q_1\neq 0$, $q_2=q_3=0$. The plot
analogous to figure \ref{fig:electric} for the electric case, is
presented in figure \ref{fig:rcharge}. The plot represents the
variation of the bare quark mass parameter $m$ as a function of the
initial boundary condition $u_0-u_H$, for Minkowski--type embeddings.
The parameter $m_*$ corresponds again to the bare quark mass for the
state corresponding to the critical embedding.  The good agreement
with the result for the critical exponents in equation
(\ref{crit-exp2}) is clear, and the accuracy of the analytic
description improves as we go deeper into the spiral (to the left).

An important observation is that our result does not depend on the
values of the R--charges, nor the temperature. In fact, this physics
persists at zero temperature, such as at extremality with all three
charges equal $q_1=q_2=q_3=q$, or more generally. (Extremality is when
$u_{E}=u_1$ or $u_2$, for which $T=0$. See
equation~(\ref{temperatureR}).) The fact that we have the same
structure at zero temperature (extremal horizon) further confirms that
the key properties of the corresponding phase transition is indeed
driven by the quantum (rather than thermal) fluctuations of the
system.

\section{Criticality and Scaling: Some Generalizations}
In this section we generalize the procedure for the study of the
critical behavior employed in all three different systems of phase
transition (thermal, or in the presence of electric field  or R--charge
chemical potential). This may lay the groundwork for other
types of phase transitions that may arise in future studies, seeded by
spirals with different universal behaviour.

Note that in all cases there is some vanishing locus. The different
classes of D$q$--brane's embeddings are being classified with respect
to whether they fall into that vanishing locus, or wrap an internal
$S^n$ sphere that is contracting to zero size above the vanishing
locus signaling the end of the D$q$--brane.

In all cases there is a critical embedding that separates the two
classes of embeddings. The critical embedding reaches the vanishing
locus and has a conical singularity there at some finite radius $u_*$
($u_* = u_H$ or $u_{\rm erg}$ for the thermal and R--charge cases). 

The main point of
the analysis is that after zooming into the space--time region near the
conical singularity we obtain the metric:
\begin{equation}
ds^2=-Dz^k dt^2+dz^2+dy^2+y^2d\Omega_n^2+\dots\ ,
\end{equation}
where $D$ is a non--essential constant.  The Dirac--Born--Infeld
Lagrangian of the brane is then:
\begin{equation}
{\cal L}\propto z^{k/2}y^n\sqrt{1+y'^2}\ .
\label{lagrgen}
\end{equation}
Note that to extract the key behavior (that we are studying) of this
critical embedding (and its neighbourhood) there is no need to modify
the embeddings which reach the vanishing locus (as we did for the
ergosphere class of embeddings) . The critical solution and the
linearized equation of motion is the same for both classes. Therefore
it is sufficient to consider the Minkowski type of embeddings and
analyze the Lagrangian (\ref{lagrgen}). The resulting equation of
motion is:
\begin{equation}
\partial_z\left(\frac{z^{k/2}y^ny'}{\sqrt{1+y'^2}}\right)-ny^{n-1}z^{k/2}\sqrt{1+y'^2}=0\ .
\label{eom-gen}
\end{equation}
It is easy to check that equation (\ref{eom-gen}) has the scaling property (\ref{scal}) and the limiting critical solution is given by:
\begin{equation}
y_*(z)=\sqrt{\frac{2n}{k}}z\ .
\label{crit-gen}
\end{equation}
Now after the substitution:
\begin{equation}
y(z)=\sqrt{\frac{2n}{k}}z+\xi(z)\ ,
\end{equation}
we obtain the following linearized equation:
\begin{equation}
z^2\xi''(z)+(n+k/2)(z\xi'(z)+\xi(z))=0\ .
\label{eom-lin-gen}
\end{equation}
The general solution of equation~(\ref{eom-lin-gen}) can be written as:
\begin{equation}
\xi(z)=\frac{1}{z^{r_n^{(k)}}}(A\cos(\alpha_n^{(k)})\ln z)+B\sin(\alpha_n^{(k)}\ln z)\ ,
\label{sol-gen}
\end{equation}
where
\begin{equation}
r_n^{(k)}=(n+k/2-1)/2;~~~\alpha_n^{(k)}=\frac{1}{2}\sqrt{4(n+k/2)-(n+k/2-1)^2};
\label{crit-exp-gen}
\end{equation}
are the scaling exponents characterizing the self--similar behavior of
the theory. Both being real, they control the shape of the spiral
which emanates from the critical solution.  The oscillatory behavior
is present for $n\leq 3+2\sqrt{2}-k/2$. For these values of $n$ the
theory exhibits a discrete self--similarity and the equation of state
$c=c(m)$ is a multi--valued function suggesting that the corresponding
phase transition is a first order one.

While there is the possibility of complex scaling exponents and hence
possibly second order phase transitions (if the multi--valuedness goes
away when the spiral does), this is not realized in the examples that
we know from string theory.

Note that we have $k=2$ for a thermal induced phase transition and
$k=1$ for the quantum induced phase transitions that we studied
(external electric field and R--charge chemical potential), arising
from the two most natural types of a vanishing locus that one may
have: an horizon, and an ergosphere. Perhaps other systems will yield
different values of $k$.

\section{Closing Remarks}

We have succeeded in casting two important types of phase transition
(in large $N_c$ gauge theory with fundamental flavours) into the same
classifying framework as the meson--melting phase transition. These
quantum fluctuation induced transitions (so--called since they persist
at zero temperature), resulting in the liberation of quarks from being
bound into mesons as a result of the application of an external
electric field, or a chemical potential for R--charge, turn out to
have the same underlying structure. It is distinct from that found for
thermal fluctuation induced transitions.  The structures are
controlled by the local geometry of the spacetime seen by a critical
D--brane embedding (it is the borderline case between two physically
distinct classes of embedding), and while it is Rindler for the
thermal case with an horizon at the origin, it is (after a T--duality
in order to geometrize the discussion as much as possible) a rotating
space with a simple ``ergosphere'' type locus.  The technique of
characterizing the physics in terms of this underlying classifying
space\cite{Frolov:2006tc,Mateos:2006nu} is rather pleasing in its
utility, and we extended our analysis to the natural generalization of
this space, extracting the scaling exponents that might pertain to
physics from future studies.

Of course, there is much interest in how much we can learn about
finite $N_c$ physics (for applications to systems such as QCD) by
studying universal features of large $N_c$. Unfortunately, it is
almost certain that much of this is far from robust against $1/N_c$
corrections. The spiral structure is rather delicate, and the stringy
corrections arising in going away from the large $N_c$ limit would
generically severely modify the classifying spacetimes we've been
studying, erasing the spiral and its self--similarity. The absence of
the spiral is necessary for there to be (at best) a second--order
transition at finite $N_c$, since it results in multi--valuedness of
the solution space, requiring the system to perform a first order
jump.

It is tempting to speculate, however, that the nature by which the
spiral is destroyed by $1/N_c$ corrections might (especially since the
setting is so geometrical) be characterizable in a way that allows
universal properties of the second (or higher) order phase transitions
to be deduced from the properties of the spiral at large $N_c$. We
leave such explorations for later work.

\section*{Acknowledgments}
We thank Tameem Albash, Nikolay Bobev, Stephan Haas, Andreas Karch,
Arnab Kundu, Krzysztof Pilch, Nick Warner, and Paolo Zanardi for
discussions.  This work was supported by the US Department of Energy.

\end{document}